# Social media in scholarly communication

Stefanie Haustein[*,1], Cassidy R. Sugimoto[2] & Vincent Larivière[1,3]

[*]stefanie.haustein@umontreal.ca
[1] École de bibliothéconomie et des sciences de l'information (EBSI), Université de Montréal, Montréal, QC (Canada)
[2] School of Informatics and Computing, Indiana University, Bloomington, IN (USA)
[3] Observatoire des Sciences et des Technologies (OST), Centre Interuniversitaire de Recherche sur la Science et la Technologie (CIRST), Université du Québec à Montréal, Montréal, QC (Canada)

## 1 Introduction

This year marks 350 years since the inaugural publications of both the *Journal des Sçavans* and the *Philosophical Transactions,* first published in 1665 and considered the birth of the peer-reviewed journal article. This form of scholarly communication has not only remained the dominant model for disseminating new knowledge (particularly for science and medicine), but has also increased substantially in volume. Derek de Solla Price—the "father of scientometrics" (Merton and Garfield, 1986, p. vii)—was the first to document the exponential increase in scientific journals and showed that "scientists have always felt themselves to be awash in a sea of the scientific literature" (Price, 1963, p. 15), as, for example, expressed at the 1948 Royal Society's Scientific Information Conference:

> Not for the first time in history, but more acutely than ever before, there was a fear that scientists would be overwhelmed, that they would be no longer able to control the vast amounts of potentially relevant material that were pouring forth from the world's presses, that science itself was under threat (Bawden and Robinson, 2008, p. 183)

One of the solutions to help scientists filter the most relevant publications and, thus, to stay current on developments in their fields during the transition from 'little science' to 'big science', was the introduction of citation indexing as a Wellsian 'World Brain' (Garfield, 1964) of scientific information:

> It is too much to expect a research worker to spend an inordinate amount of time searching for the bibliographic descendants of antecedent papers. It would not be excessive to demand that the thorough scholar check all papers that have cited or criticized such papers, if they could be located quickly. The citation index makes this check practicable. (Garfield, 1955, p. 108)

In retrospective, citation indexing can be perceived as a pre-social web version of *crowdsourcing,* as it is based on the concept that the community of citing authors outperforms indexers in highlighting cognitive links between papers, particularly on the level of specific ideas and concepts (Garfield, 1983). Over the last 50 years, citation analysis and, more generally, bibliometric methods, have developed from information retrieval tools to research evaluation metrics, where they are presumed to make scientific funding more efficient and effective (Moed, 2006). However, the dominance of bibliometric indicators in research evaluation has also led to significant goal displacement (Merton, 1957) and the oversimplification of notions of "research productivity" and "scientific quality", creating adverse effects such as salami publishing, honorary authorships, citation cartels, and misuse of indicators (Binswanger, 2015; Cronin and Sugimoto, 2014; Frey and Osterloh, 2006; Haustein and Larivière, 2015; Weingart, 2005).

Furthermore, the rise of the web, and subsequently, the social web, has challenged the quasi-monopolistic status of the journal as the main form of scholarly communication and citation indices as the primary assessment mechanisms. Scientific communication is becoming more open, transparent, and diverse: publications are increasingly open access; manuscripts, presentations, code and data are shared online; research ideas and results are discussed and criticized openly on blogs; and new peer review experiments, with open post publication assessment by anonymous or non-anonymous referees, are underway. The diversification of scholarly production and assessment, paired with the increasing speed of the communication process, leads to an increased information overload (Bawden and Robinson, 2008), demanding new filters.

The concept of *altmetrics*, short for alternative (to citation) metrics, was created out of an attempt to provide a filter (Priem et al., 2010) and to steer against the oversimplification of the measurement of scientific success solely on the basis of number of journal articles published and citations received, by considering a wider range of research outputs and metrics (Piwowar, 2013). Although the term *altmetrics* was introduced in a tweet in 2010 (Priem, 2010), the idea of capturing traces —"polymorphous mentioning" (Cronin *et al.*, 1998, p. 1320)—of scholars and their documents on the web to measure "impact" of science in a broader manner than citations was introduced years before, largely in the context of webometrics (Almind and Ingwersen, 1997; Thelwall et al., 2005):

> There will soon be a critical mass of web-based digital objects and usage statistics on which to model scholars' communication behaviors—publishing, posting, blogging, scanning, reading, downloading, glossing, linking, citing, recommending, acknowledging—and with which to track their scholarly influence and impact, broadly conceived and broadly felt. (Cronin, 2005, p. 196)

A decade after Cronin's prediction and five years after the coining of *altmetrics*, the time seems ripe to reflect upon the role of social media in scholarly communication. This special issue does so by providing an overview of current research on the indicators and metrics grouped under the umbrella term of *altmetrics*, on their relationships with traditional indicators of scientific activity, and on the uses that are made of the various social media platforms—on which these indicators are based—by scientists of various disciplines.

## 2 Terminology and Definition
The set of metrics commonly referred to as *altmetrics* are usually based on the measurement of online activity related to scholars or scholarly content derived from social media and web 2.0 platforms. As such, they can be considered as a proper subset of webometrics and scientometrics. However, the definition of what constitutes an "altmetric" indicator is in constant flux, as it is largely determined by technical possibilities and, more specifically, the availability of application programming interfaces (APIs). The common denominator of various *altmetrics* is that they *exclude* and stand opposed to 'traditional' bibliometric indicators (see e.g., Priem *et al.* (2010)), and often include usage metrics—despite the fact that these indicators have been available much longer and are not based on social media platforms (Haustein, 2014). More recently, and quite inclusively, Priem (2014, p. 266) defined the field of altmetrics as the "study and use of scholarly impact measures based on activity in online tools and environments".

*2.1 The name debate: altmetrics, article level metrics, social media metrics – or just metrics?*
There has been considerable debate and confusion surrounding the meaning of the term *altmetrics*. Many have seen PLOS's article level metrics (ALM) program (Fenner, 2013)—the first major attempt to systematically provide numbers on papers' bookmarks on CiteULike and Connotea; mentions on blog posts, reader comments, and ratings; as well as citations, article views, and downloads—as synonymous with altmetrics. However, a criticism of article level metrics as being too constraining was bound up in the origin of the term *altmetrics* (see tweet by Jason Priem, Figure 1). As discussed above, Priem later (2014) broadened the definition to include scholarly impact measures available on any online platform.

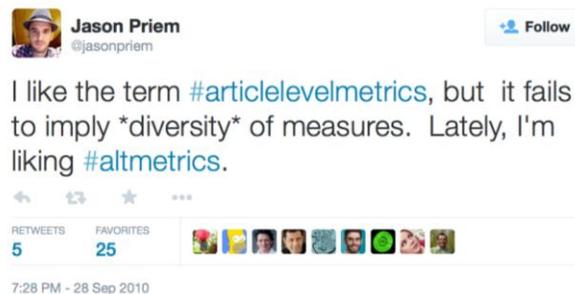

**Figure 1**. The tweet by Jason Priem, which coined the term *altmetrics*.
https://twitter.com/jasonpriem/status/25844968813

The scientometric community quickly responded. Rousseau and Ye (2013, p. 2), claimed that altmetrics was "a good idea but a bad name" and proposed to replace it by *influmetrics,* which "suggest[s] diffuse and often imperceptible traces of scholarly influence – to capture the opportunities for measurement and evaluation afforded by the new environment" (Cronin, 2005, p. 176). The term was introduced by Davenport and initially discussed by Cronin and Weaver (1995) in the context of acknowledgements and webometrics. Emphasizing the origin of the data instead of intent or meaning of the new metrics, Haustein *et al.* (2014b) proposed the term *social media metrics*. However, with the changing landscape of platforms and definitions shaped by data collection methods of aggregators and vendors, the term social media metrics might be too restrictive. For example, Plum Analytics incorporated library holdings (Parkhill, 2013) and Altmetric.com monitors newspapers and have started to include mentions in policy documents (Liu, 2014). Thus, the heterogeneity and dynamicity of the scholarly communication landscape make a suitable umbrella term elusive. It may be time to stop labeling these terms as parallel and oppositional (i.e., altmetrics vs bibliometrics) and instead think of all of them as available scholarly *metrics*—with varying validity depending on context and function.

*2.2 In search of meaning: interpreting and classifying various metrics*
Data aggregators and providers like PLOS and ImpactStory were the first to categorize the types of impact based on data sources. PLOS, for instance, categorizes data sources as *viewed*, *saved*, *discussed, cited* and *recommended*, assuming increasing engagement from viewing to recommending (Lin and Fenner, 2013). ImpactStory uses the same categories, but distinguishes between scholars and the public as two distinct audiences (Piwowar, 2012). However, these platform-based distinctions are quite general and, often, based on what they *intend* to measure rather than what is *actually* measured by the category of indicator. For example, ImpactStory categorizes HTML views as views by the public, while PDF downloads are considered as being made by scholars. Similarly, the platform considers tweets as being made by the general public, although many tweets associated with scientific papers are likely to come from researchers (Tsou *et al.*, in press). More recently, social media metrics have been discussed in light of citation theories (i.e., normative, social constructivist approaches and concept symbols) and social theories (i.e., social capital, attention economics and impression management) to contribute to the understanding of the meaning of the various metrics (Haustein *et al.*, in press). This has led to a framework that categorizes various acts related to research objects (i.e., scholarly documents and agents) into the three categories of *access*, *appraise* and *apply*, rather than classifying the indicators based on the tools and data sources from which they come.

**3 Current research**
Since the coining of the term *altmetrics* in 2010, there has been a proliferation of scholarship on the subject of social media and scholarly communication. We provide here a brief overview of the current research related to the use and role of social media in scholarly communication, as well as the metrics derived from this use.

*3.1 Social media uptake and motivation in academia*
A number of social media tools and platforms have been developed to allow for the dissemination and access of scholarship, the communication and interaction among scholars, and the presentation of profiles at various levels of aggregation (e.g., individual, journal, institution). Persistent questions in social media metrics have been the extent to which the platforms are used, why, and by whom—critical questions for appropriate generalization and decision-making on the basis of these platforms.

High degrees of use of social media and networking tools have been demonstrated at the individual level with percentages as high as 75% (Tenopir *et al.*, 2013) and 80% (Procter *et al.*, 2010), although uptake varies among fields and by demographic characteristics (e.g., gender, age). However, the surveys reporting these numbers were highly inclusive—operationalizing social media and networking tools to include Skype and Wikipedia. Looking more closely at particular social media platforms shows substantial variation, with Google Scholar (Haustein *et al.*, 2014c; Procter *et al.*, 2010), collaborative authoring tools (Rowlands *et al.*, 2011), and LinkedIn (Haustein *et al.*, 2014c; Mas-Bleda *et al.*, 2014) among the most popular. Lower rates have been found for other social media sites: rates of Twitter use for academics is around 10% (Grande *et al.*, 2014; Procter *et al.*, 2010; Pscheida *et al.*, 2013; Rowlands *et al.*, 2011), trailed by Mendeley (6%), Slideshare (4%), and Academia.edu (2%) (Mas-Bleda *et al.*, 2014). However, many of the studies have been disciplinarily homogeneous and have shown extreme variation based on the population (e.g., the rates of tweeting among bibliometricians in Haustein *et al.* (2014c) or the rates of blogging among academic health policy researchers in Grande *et al.* (2014)).

Individual motivations to use social media for scholarly communication vary significantly by country (Mou, 2014; Nicholas *et al.*, 2014), age (Nicholas *et al.*, 2014), and across and within platforms (Mohammadi *et al.,* in press b). While some have touted the advantages for collaboration and the age-bias of social media, others have challenged these claims (Harley *et al.*, 2010). The demographics of those who employ these technologies is also variable—e.g., Mendeley has been shown to be dominated by graduate students (Mohammadi *et al.*, in press a; Zahedi *et al.,* 2013). Imbalances in terms of gender (Shema *et al.*, 2012), level of education (Kovic *et al.*, 2008), and disciplinary area (Shema *et al.*, 2012) also call for cautious interpretations of the analyses derived from a single platform.

Academic institutions have implemented social media tools to varying degrees. Motivations for institutional use of social media ranges from faculty development (Cahn *et al.*, 2013) to pedagogy (Kalashyan *et al.*, 2013). Academic libraries have been early adopters of social media tools, with nearly all libraries maintaining an institutional Twitter and Facebook account as well as hosting a blog (Boateng and Quan Liu, 2014). Journals have also increasingly adopted social media tools, using commenting (Stewart *et al.*, 2013), blogging (Kortelainen and Katvala, 2012; Stewart *et al.*, 2013), and social networking (Kortelainen and Katvala, 2012).

*3.2 Analysis of social media metrics*
The majority of published studies on the topic have focused on social media activity associated with journal articles. Most of these examine the extent to which scientific articles are visible on various platforms (coverage), the average attention they receive (mean event rate), and the degree to which the metrics correlate with citations and other metrics. In terms of signal, Mendeley (the social bookmarking platform) has been shown to be the dominant source, with levels of coverage as high as 50-70% in some disciplines (e.g., biomedical research and the social sciences) and nearly ubiquitous coverage for some journals (e.g., *Nature*, *Science*, *JASIST*, and *PLOS* journals) (Haustein *et al.*, 2014b; Bar-Ilan, 2012; Li *et al.*, 2012; Mohammadi *et al.*, in press a; Mohammadi and Thelwall, 2014; Priem *et al.*, 2012). Other social reference managers such as CiteULike and BibSonomy capture less activity (Haustein and Siebenlist, 2011; Li *et al.*, 2012); for example, 31% of *PLOS* articles were bookmarked on CiteULike compared to 80% on Mendeley (Priem *et al.*, 2012). As shown with other metrics, there are certainly country-affiliation advantages (Sud and Thelwall, in press).

Coverage and mean event rates for Twitter have been shown to be lower than Mendeley— between 10%-21%, depending on the study and corpus (Costas *et al.*, in press; Haustein *et al.,* 2015; Priem *et al.*, 2012). There are also significant differences in across fields (Haustein *et al.*, 2014b) and subfields (Haustein *et al.*, 2014e). Facebook has even lower rates of coverage (between 2.5-10%, depending on study), though the access to these events is limited to publicly available profiles and thus has high potential for missing data (Costas *et al.*, in press; Haustein *et al.,* 2015; Priem *et al.*, 2012). Similarly, access to comprehensive data on the mention of articles in blogs has proven to be difficult. Current studies estimate between 2-8% coverage for this media (Costas *et al.*, in press; Haustein *et al.,* 2015; Priem *et al.*, 2012), varying by discipline, journal, and open access policy (Fausto *et al.*, 2012; Groth and Gurney, 2010). However, these are early years for social media metrics. Additional platforms are being developed and existing platforms are now recognized for their potential to inform scholarly assessment: for example, Goodreads (Zuccala *et al.*, 2014; Zuccala *et al.*, in this issue) and Wikipedia (Evans and Krauthammer, 2011; Nielsen, 2007; Priem *et al.*, 2012). One fairly recent advance has been the metricization of peer review, through systems such as F1000. While few studies have sought to examine coverage (with the exception of Priem *et al.*, 2012), many authors have explored the various recommendation categories and levels (Waltman and Costas, 2014), disciplinary representation (Waltman and Costas, 2014), and correlations between these categories and other metrics (Bornmann in this issue; Waltman and Costas, 2014).

One constant has been the assumption that the validity and utility of new metrics can be tested through correlational analyses with traditional bibliometrics indicators (Li *et al*, 2012). The majority of results have reported mostly weak correlations between citations and various social media metrics (Bornmann and Leydesdorff, 2013; Costas *et al.*, in press; Eysenbach, 2011; Fausto *et al.* 2012; Haustein *et al.,* 2014b, 2014d, 2015; Mohammadi and Thelwall, 2013; Thelwall *et al.*, 2013), though some have found moderately strong positive relations (Haustein *et al.*, 2014b; Nielsen, 2007). As with all metrics, strong variation is seen by the population under analysis (e.g., Shema and Bar-Ilan (2014)), making it difficult to generalize the results. Correlational analyses have also examined the relation among social media metrics—for example, between downloads and reference manager saves (Priem *et al.*, 2012), tweets and downloads (Shuai *et al.*, 2012), F1000 metrics and social media metrics (Bornmann, in this issue; Li and Thelwall, 2012), blog posts and social media metrics (Allen *et al.*, 2013), and F1000 metrics and expert assessment (Allen *et al.*, 2009). The interpretation is always difficult—in case where correlations are positive and significant one questions whether the new metric is duplicative and therefore unnecessary; insignificant correlations may signal that something distinct has been measured (e.g., Sugimoto *et al.* (2008)). A third option, significant negative correlations—as shown in Haustein *et al.* (2014f)—may be the strongest prediction of distinction among the measures.

*3.3 Data reliability and validity*
There has been considerable concern over the reliability and validity of social media metrics (e.g., Dinsmore *et al.* (2014); Nature Materials Editors (2012)). In a comprehensive survey of more than 15 tools used to generate social media metrics, Wouters and Costas (2012, p. 5) concluded that altmetrics need a "far stricter protocol of data quality and indicator reliability and validity" before they could be appropriately applied to impact assessment. Many of the concerns regard data collection techniques and the variability among sources and time of collection (Gunn, 2014; Neylon, 2014; Torres-Salinas *et al.*, 2013), which affects replicability of the research. Methodological and statistical concerns are also paramount—there is a need to codify standard practices to the analysis of social media metrics (e.g., Sud and Thelwall, (2013)).

Furthermore, the validity of these measures is called into question, given that the most tweeted papers, for example, often have funny titles, report curious topics (Haustein *et al.*, 2014d), and refer to "the usual trilogy of sex, drugs, and rock and roll" (Neylon, 2014, para. 6). Social media metrics are often seen as positive indicators of public interest in science; however, these results are complicated by the lack of knowledge about the demographics of those utilizing the platforms and the presence of automated profiles (or bots) engaging in the system (Haustein *et al.*, 2014a). Perhaps the most important criticism is the degree to which the focus on and proliferation of new metrics causes a displacement of attention from scholarship to social media performance (Gruber, 2014).

## 4 Contribution of this special issue

Out of the 22 submissions received, 6 papers were accepted, for an acceptance rate of 27%. The submitted contributions were reviewed by 37 external reviewers. The six accepted manuscripts are complementary to each other and fill some of the gaps currently found in the literature.

The first paper of the issue, authored by Rodrigo Costas, Zohreh Zahedi, and Paul Wouters from the Center for Science and Technology Studies (CWTS) of Leiden University, uses science maps to compare the visibility of papers on various social media platforms with scores obtained using traditional bibliometric indicators. Drawing on more than half a million papers published in 2011, they visualize which subjects areas (as presented in the map of science produced by CWTS) are popular on Twitter, Mendeley, Facebook, blogs, and mainstream news as captured by Altmetric.com. They highlight the similarity between citations and Mendeley readership in terms of the research areas in which the counts are most frequent, and show that, for most disciplines, readership counts exceed citation rates. This was especially true for the social sciences. They authors also show that papers in general medicine, psychology, and social sciences—fields that are considered to have greater social impact—are much more visible on Twitter than papers in other fields, which suggests that tweets could, to a certain extent, reflect impact on the general public. Mentions on less prevalent platforms such as Google+, blogs, and mainstream media show biases towards papers published in multidisciplinary journals such as *Nature*, *Science,* or *PNAS*. Regarding Mendeley, Costas, Zahedi, and Wouters conclude that, in the social sciences (much more than in the humanities and natural sciences), where the use of citations is more problematic, readership counts could be used as an alternative to citations as a marker of scientific impact.

Also using Altmetric.com data, Juan Pablo Alperin (Simon Fraser University) tackles another important issue in science indicators: the geographical bias of altmetrics. Using the metadata of papers indexed in the Latin American journal portal SciELO—which indexes more than 1,200 journals and half a million articles—he measures the coverage (i.e., proportion of articles with non-zero values) of Mendeley, Facebook, Twitter, and other metrics provided by Altmetric.com across the different disciplines and compares the results with those obtained in other studies that used "international" databases. He shows that papers indexed on SciELO obtain lower coverage than those of papers indexed in other databases, with scores close to zero in most cases. This was also true for the major Brazilian collection—the largest in SciELO. Alperin suggests three potential explanations for this: 1) SciELO has a lower usage and, thus, a lower social media usage; 2) social media use is lower in Latin America than elsewhere in the previously studied contexts; or 3) Latin American has distinct practices of sharing research on social media. In sum, Alperin's results convincingly demonstrate that, in addition to discipline and topicality of papers, geography affects the visibility of papers on social media platforms.

The next paper, by Lutz Bornmann of the Max Planck Society, focuses on relationships among a subset of altmetrics (namely tweets and Facebook scores) with various tags assigned by experts to papers on F1000, using a sample of 1,082 papers published in PLOS journals. Counts on Facebook and Twitter were significantly higher for papers tagged on F1000 as "good for teaching" than for papers without this tag. Bornmann also observes that the number of Mendeley readership counts is positively associated with the use of the tag "technical advance", which is assigned to papers that are considered by experts as introducing a new practical or theoretical technique, and that the "new finding" tag is positively related with the number of Facebook posts. Using the tag "good for teaching" as a marker of the potential impact of a paper beyond specialized researchers of a discipline, Bornmann argues that Twitter and Facebook counts, but not those from Figshare or Mendeley, might be useful for measuring the social impact of research.

The paper by Alesia Zuccala, Frederik Verleysen, Roberto Cornacchia and Tim Engels (University of Copenhagen, University of Antwerp and Spinque B.V.) analyses the usefulness of an original data source for informetric research, Goodreads (a social cataloguing platform on which readers can rate and recommend books), for measuring the wider impact of academic books. Drawing on books cited by 604 history journals from the Scopus database, the authors retrieved a list of more than 8,500 history books from the Goodreads platform, for which they compared the citation and reader rating counts. For the entire dataset of books as well as the subset that received both a high number of citations and reader ratings, low correlations were found between citations and reader rating counts, which suggest that Goodreads ratings could be used as a complement to citations. Their results also shows that reader ratings were more likely to be given to books held in academic and public libraries outside the United States, which suggests a positive effect of the books' international visibility. Of course, as with any new data source, more research is needed to assess whether these findings can be obtained using other datasets covering different disciplines; however, their method provides a unique window on assessing the impact of research in a domain—history—that has remained for several decades one of the blind spots of bibliometrics and research evaluation.

Focusing on visualizing and interpreting social media activity, Victoria Uren and Aba-Sah Dadzie (Aston University and University of Birmingham), compare the Twitter activity of a trending—if not viral—topic (the Curiosity landing) with that of two non-trending topics (Phosphorus and Permafrost), to assess whether methods used for the first group of topics could be transferred to the second group of topics. Results show that the parallel coordinates visualization method, in combination with pattern matching, is an effective method for observing dynamic changes in Twitter activity, as it allows for the analysis of both midsize and large collections of microposts, and provides the scalability required for longitudinal studies. As a large proportion of the research on altmetrics has been performed by information scientists who have transposed methods and frameworks from the bibliometric paradigm to the analysis of altmetrics, this original methodological contribution provides researchers in the field with more advanced methods to study the diffusion of scientific information on the microblogging platform. The method differs from standard approaches to the analysis of microblog data, which tend to focus on machine driven analysis of large-scale datasets. It provides evidence that this approach enables practical and effective analysis of the content of midsize to large collections of microposts.

The final paper in this issue also examines Twitter. Timothy D. Bowman (Université de Montréal and Indiana University Bloomington) provides the results of an analysis of tweeting behaviors by professors affiliated to universities of the Association of American Universities (AAU), based on both a survey of the faculty as well as an in-depth analysis of their tweets. A random sample of 75,000 of professors' tweets were classified as personal or professional based on the impression of users on Amazon Mechanical Turk, so-called 'turkers'. Bowman's findings emphasize the differences in various disciplines' usage of Twitter, with half of the computer scientists surveyed having an account, compared to about one-fifth of chemists. Younger researchers were also more likely to have an account than older researchers. In all departments surveyed, respondents indicated using Twitter in both personal and professional contexts, or for professional reasons only, except in philosophy where most professors used it for strictly personal reasons. Differences were also found in the use of affordances—such as #hashtags, @user mentions, URLs, and retweets—across personal and professional tweets (as classified by turkers), which suggests that different social norms frame the various uses of the platform. Bowman also shows that tweeting activity (i.e., number of tweets per day) varied greatly across disciplines, with scholars from the social sciences tweeting more (1.40 per day) than scholars from the natural sciences (0.61 per day). As one of the largest analyses of the prevalence of Twitter use in academia and of the various usages and factors that influence its uses, this paper contributes to the development of a theoretical framework that allows for the interpretation of Twitter-based indicators of science.

## 5 Conclusion

The contributions in this special issue provide insights on social media activity related to scholars and scholarly content, as well as on the metrics that are based on these online events. After decades of studying scholarly communication almost exclusively with papers and citations, scholars now have access to new sources of evidence which, in turn, has brought new energy to the science indicators community.

Several parallels can be drawn between the current state of research on social media metrics and the early days of citation analysis. In a manner similar to the *altmetrics* research community, the bibliometric community has historically been driven by data availability rather than by crafting indicators based on specific concepts. In that sense, both communities (which overlap to a certain extent) have been quite pragmatic. However, while citations had been a central and established component of scholarly communication since the early days of modern science, the role and uses of various social media platforms within and outside academe are still taking shape (Haustein *et al.*, in press). At the same time, funders, universities, and publishers increasingly demand indicators of the impact of science on society.

The comparison with bibliometrics can also provide us with lessons learned, as researchers are increasingly observing the adverse effects of the use of such indicators in research evaluation (Binswanger, 2015; Frey and Osterloh, 2006). Let us not condemn the burgeoning field of altmetrics to the same fate. As altmetrics hold the potential to make the evaluation of research activities more comprehensive, we need to focus our attention on understanding the meaning of these metrics. Hopefully, this special issue is a step in this direction.

Guest Editorial *Aslib Journal of Information Management* 67(3)
Special Issue "Social media metrics in scholarly communication: exploring tweets, blogs, likes and other altmetrics"

**Acknowledgements**
We would like to thank all authors for submitting their manuscript and contributing to this special issue as well as the 37 reviewers for their valuable feedback. We also thank Sam Work for her help with the literature review and acknowledge funding from the Alfred P. Sloan Foundation Grant #G-2014–3–25.